\documentclass[
 reprint,
 amsmath,amssymb,
 aps,prb,eps,superscriptaddress,twocolumn,10pt
]{revtex4}
\usepackage{gensymb}
\usepackage{graphicx}
\usepackage{dcolumn}
\usepackage{bm}
\usepackage{xcolor}
\usepackage{hyperref}
\newcommand{\HB}[1]{\textcolor{black}{#1}}
\newcommand{\MA}[1]{\textcolor{black}{#1}}

\begin{document}

\preprint{APS/123-QED}
\title{Temperature and Pressure driven Spin transitions and Piezochromism in a Mn-based Hybrid Perovskite}

\author{Hrishit Banerjee}
 \email{hb595@cam.ac.uk}
 \affiliation{Department of Chemistry, University of Cambridge, Lensfield Road, Cambridge, CB2 1EW, UK}
 \affiliation{Institute of Theoretical and Computational Physics, Graz University of Technology, NAWI Graz, Petersga{\ss}e 16, Graz, 8010, Austria.}
\author{Angela Rittsteuer}
\author{Markus Aichhorn}%

\affiliation{Institute of Theoretical and Computational Physics, Graz University of Technology, NAWI Graz, Petersga{\ss}e 16, Graz, 8010, Austria.}

\date{\today}

\begin{abstract}
Hybrid perovskites have been at the forefront of condensed matter research particularly in context of device applications primarily in relation to applications in the field of solar cells. In this article, we demonstrate that several new functionalities may be added to the arsenal of hybrid perovskites, in terms of external stimuli driven spin transitions as well as piezochromism. As an example, we study Dimethylammonium Manganese Formate (DMAMnF), a hybrid perovskite investigated quite extensively experimentally. We show by employing first principles DFT+U calculations with the aid of ab initio molecular dynamics calculations that
DMAMnF shows temperature and pressure driven spin transitions, from a low spin $S=1/2$ to a high spin $S=5/2$ state. This transition is accompanied by a hysteresis, and
we find that this hysteresis and the transition temperature are quite close to room temperature, which is desirable for device applications particularly in memory, display, and switching devices. The operating pressure is a few GPa, which is 
accessible in standard laboratory settings. We find that the cooperative behaviour showing up as hysteresis accompanying the transition is driven primarily by elastic interactions, assisted by magnetic superexchange between Mn atoms. 
Last but not least we demonstrate that the spin transition is associated with piezochromism which 
could also be important for applications.
\end{abstract}

\maketitle

\section{Introduction}

In recent times, dense metal-organic hybrid frameworks with low porosity have garnered considerable attention from the communities involved in both solid state chemistry and materials research \cite{metal-org1, metal-org2, stuart, cote, sensors, catalysis, biomedical, morris, kreno}. Some of these dense hybrid compounds adopt the famous perovskite geometry with the general formula ABX$_3$. This has opened up an emerging area of research on hybrid organic-inorganic perovskites, parallel to the well established field of inorganic perovskite oxides. Hybrid perovskites are, thus, organic counterparts of inorganic perovskites. In case of hybrid perovskites, the B cation is a metal ion as in inorganic perovskites, but both the A cation as well as the X ligand can be organic. The family of lead halide hybrid perovskite having [AmH]MX$_3$ composition, where AmH$^+$ is the protonated amine part, M is either Sn$^{2+}$ or Pb$^{2+}$, and X is the halogen part (Cl$^-$, Br$^-$, or I$^-$) have been shown to demonstrate high performance and efficiency in applications related to the design of mesostructured and nanostructured solar cells and other photovoltaic devices, which are particularly important for energy based applications \cite{Prasanna2019, Manser2016, Kojima2009, Cui2015, Ching, Holovsky2019}. Graphene-enabled hybrid perovskite solar cells are already in use in real life applications \cite{BOUCLE20163}. Easy processing techniques like spin-coating, dip-coating, and vapor-deposition techniques have been known to be of advantage in this case. 

There are also transition-metal-formate based hybrid perovskites, which are of interest in the current context, having formula [AmH]M(HCOO)$_3$ (M = Mn, Cu, Ni, Fe, Co)~\cite{hybrid-formate}. In the crystal structure of these compounds, formate bridges act as linkers that connect the MO$_6$ octahedra, with the protonated amine cations situated at the hollow spaces formed by the linked octahedra. These hollow spaces act as pseudo-cubic ReO$_3$ type cavities. Organic ligands like formate being simple enough have been mostly studied with varied A cations. This class of materials has been shown to exhibit curious properties, of which multiferroicity seems to be an intriguing one~\cite{Jain1, Jain2}. Ferroelectricity and especially multiferroicity in these materials has been extensively studied by Stroppa and coworkers mostly from a density-functional theory (DFT) based first-principle perspective, at times combined with experimental studies~\cite{stroppa1,stroppa2, stroppa4, stroppa5, Jain2016, sante, Tian2014, aguirre, kamminga, ptak, ghosh, mazzuca}. \HB{Order-disorder phase transitions as a function of temperature and pressure have also been well studied~\cite{Clune2020, Chitnis2018, Sanchez-Andujar2010}.} Structural details and effects due to structural phase transitions, strain tuning of various effects like polarisation, and the magnetic structure has also been studied by the same groups. The presence of organic components in the structure offers significant structural flexibility and thus enables a better tunability of properties by external means. The flexibility of hybrid perovskites to undergo large structural changes in response to external stimuli has been already reported.

The presence of transition metal ions in these compounds together with its octahedral environment makes them well suited for exhibiting spin transitions. Spin transition behavior under the influence of external stimuli like temperature, pressure~\cite{Saha-Dasgupta2014, sco2, sensors, sudipto-sco1, sudipto-sco2} or light irradiation~\cite{OLGUIN2011}, and possibly also cooperativity resulting in hysteresis~\cite{Kahn1, Kahn44, Banerjee2014, Banerjee2016, Banerjee2017, Boukheddaden}, have been well studied in metal-organic coordination polymers. The demonstration of cooperative spin transitions in hybrid perovskites, which are made possible by the dense nature of the framework, would add another dimension to the functionality of this interesting class of materials. However, the aspect of spin transitions in this interesting class of compounds has remained largely unexplored, apart from very recent theoretical proposals, albeit on a theoretically predicted Fe-based material~\cite{Banerjee2016, Banerjee2017}.  Another intriguing possibility is the change in colour in similar hybrid perovskite materials, bringing in the idea of piezochromism. This has recently been explored in a series of lead-based \cite{Umeyama2016} as well as Cu-based \cite{Jaffe2015} and lead-free alkaline-earth chalcogenide perovskites~\cite{arnab-piezo}. 

In this study, we focus on these two particularly less explored areas, namely, the external stimuli-driven spin transition and piezochromism in hybrid perovskites from a first-principles perspective. A very desirable property in the context of spin transitions from the point of view of device applications is cooperativity in the spin transition phenomena, which usually manifests itself by an associated hysteresis. This has important implications in designing memory devices~\cite{Saha-Dasgupta2014, Kahn1, Kahn44, Banerjee2014, Banerjee2016, Banerjee2017}. Spin transitions without hysteresis also finds application in optical switches, sensors, and alike~\cite{sensors}. For device applications, the prime concerns are (i) to achieve large hysteresis width, which would enable the memory effect to be observable over a wide range of external stimuli, and (ii) the transition to occur at a value of the stimuli that can be reached readily, for example small values of pressure or room temperature conditions. Considering the dense network structure of the discussed hybrid perovskites, they can potentially exhibit cooperative spin-state transitions, thus adding a new functionality to the already large arsenal in this very interesting class of materials. Piezochromism as well has several application possibilities in the design of materials and coatings. Those devices, which are sensitive to external perturbations, can find varied use in detectors, displays, etc.

Although most of the studies on spin transitions have focused primarily on Fe-based metal-organic complexes, there exist as well some studies on Co~\cite{Taylor, Schweinfurth2014, KRIVOKAPIC2007364}, Mn~\cite{Amabilino2017, kazakova, magnetochemistry2010001, Chen2014sco, Alcover}, or Ni~\cite{Qamar, Homma2018, Ma2011} based metal-organic compounds. Here, we consider a Mn-based hybrid perovskite. The choice is prompted by the fact that this compound exists in its ordered state and has been synthesized and studied experimentally for various properties. It has five $d$-electrons in the Mn(II) ion and shows a pronounced and abrubt transition from a low-spin (LS) $S = 1/2$ to a high-spin (HS) $S = 5/2$ state, which makes these materials suitable for applications. In our study, we consider both temperature and hydrostatic pressure as external stimuli, as well as compare the absorption spectra in the different spin states and at different pressures to study the phenomenon of piezochromism. To the best of our knowledge, no comprehensive study exists so far on all these external stimuli effects together on an experimentally available formate-based hybrid perovskite. 

Our first-principles computational study is based on DFT, supplemented with a Hubbard $U$ approach (DFT+U), along with ab-initio 
molecular dynamics (AIMD) simulations. It takes into account all structural and electronic aspects, and their responses to external stimuli in full rigor, and it shows that both temperature and pressure-induced spin-state transitions are achieved in dimethylammonium manganese formate (DMAMnF) for reasonably experimentally accessible critical values. For the pressure-driven transition, we find a critical value of $\sim 5$\,GPa, associated with a large hysteresis of $\sim 2$\,GPa. In case of temperature as external stimulus, the transition temperature is of the order of $\sim 300$\,K with a hysteresis width of $\sim 50$\,K, taking the spin-transition phenomena to room temperatures. Both the large hysteresis width in case of pressure and temperature driven spin transitions, as well as having the transitions in an accessible range (moderate operating pressure and room temperature) are highly desirable for practical device applications. Our studies also show very different absorption spectra in the LS and the HS state, confirming a change in colour between the two spin states, 
holding great promise for piezochromism-based applications as well. This is to the best of our knowledge the first demonstration of piezochromism associated with spin state transitions. Modification of the characteristics of the optical absorption spectra along with reasonable red shifts upon compression have been reported~\cite{Jaffe2015, Umeyama2016, arnab-piezo}.

Our results show that these compounds exhibit spin-switching properties over a wide range of moderate operating pressure and room temperature conditions. \MA{Although the necessary pressures are still far too large for practical device applications, our study shows a pathway of how these materials could be used in principle in such settings.} 
The piezochromism observed with applied pressure may have significant impact in understanding the optical response of this class of materials \MA{also} from a fundamental point of view. 

\section{Computational Details}
Our first-principles calculations were carried out in the plane-wave basis as implemented in the Vienna Ab-initio Simulation Package (VASP) \cite{kresse, kresse01} with projector-augmented wave (PAW) potentials~\cite{blochl}. As exchange-correlation functional we use the generalized gradient approximation (GGA) implemented following the Perdew–Burke–Ernzerhof prescription~\cite{pbe}. For ionic relaxations, internal positions of the atoms were relaxed until the forces became less than 0.005\,eV/Å. A plane-wave energy cutoff of 550\,eV and a 6$\times$6$\times$4 Monkhorst-Pack k-points mesh were found to provide a good convergence of the total energy in self-consistent calculations. To take into account the correlation effect at Mn sites beyond GGA, which turned out to be crucial for the correct description of the electronic and magnetic properties, we performed DFT+U calculations as suggested by Liechtenstein \textit{et al.}~\cite{ldau}, where we use a Hubbard interaction parameter of $U = 3.5$\,eV and Hund’s coupling parameter $J = 0.9$\,eV. In order to study the effect of hydrostatic pressure, calculations were done by first changing the volume of the unit cell isotropically and then relaxing the shape of the cell together with the ionic positions for each of the modified volumes. 
Accurate self-consistent-field calculations were carried out to obtain the total energy of the systems at each volume. The energy versus volume data was fitted to the third order Birch–Murnaghan isothermal equation of state \cite{Murnaghan1944},
given by,
\begin{align}
 E(V) = E_0+\frac{9V_0B_0}{16}\lbrace \left[(\frac{V_0}{V})^{2/3}-1\right]^3 B'_0 + & \notag  \\ \left[(\frac{V_0}{V})^{2/3}-1\right]^2 \left[6-4(\frac{V_0}{V})^{2/3}\right] \rbrace 
\end{align}
where $V_{0}$ is the equilibrium volume, $B_{0}$ is the bulk modulus at $V_0$, and $B'_{0}$ is the
pressure derivative of $B_{0}$. The bulk modulus connects the volume and pressure changes as 
\begin{equation}
B_{0}=-V\frac{\delta P}{\delta V}\Big|_{T}
\end{equation}
which we use to calculate an estimate for the applied hydrostatic pressure from the volume changes.

In order to investigate the temperature driven transition in the above-mentioned hybrid perovskites, we carried out ab-initio molecular dynamics (AIMD) calculations as implemented in VASP~\cite{hafner}. We carry out thermalisations using the NpT and consequently the NVE ensembles. For this purpose, the LS structure was initially optimized using DFT at 0\,K. The positions of all unconstrained atoms were relaxed until the forces became less than 0.005\,eV/\AA, 
and the change in bond lengths less than $10^{-3}$\,\AA. Starting from the $T=0$\,K optimized structures, the temperature was increased in steps of 5\,K using a Berendsen thermostat~\cite{berendsen} with a time step of 1\,fs for each molecular dynamics step. Close to the transitions we did even denser temperature steps of 1\,K.
At the final temperature, the system was thermalised for a time duration of 1\,ps using the NpT ensemble and the Rahman-Parinello algorithm~\cite{rahman-parrinello1, rahman-parrinello2}, using the Langevin thermostat~\cite{Allen2017} to account for the change in volume at that temperature. At the end, a NVE ensemble thermalization was carried out to accurately determine the energy of the system at the final temperature, and the average temperature was recorded, after fluctuations in temperature reduced below a certain cutoff. \HB{This in effect also takes into account all the possible enthalpy contributions.} It is to be noted that there is not really a standard prescription for carrying out temperature evolution of a system. We try to emulate here the actual process of heating a system slowly in a experiment. 

\section{Results}
\subsection{Crystal structure}
An essential prerequisite for a first-principles study is the accurate information on the crystal structure, so we start in this section with a brief description of the structure of DMAMnF. An interesting feature of [AmH]M(HCOO)$_3$ compounds is the order-disorder transition \cite{Clune2020, Sanchez-Andujar2010} of the A-site amine cations through ordering of hydrogen bonds. Crystal structure data for both the ordered and disordered phases of DMAMnF are available on ICSD. We start with the ordered phase crystal structure, which has $Cc$ space group \HB{[ICSD 261977 \cite{Sanchez-Andujar2010}]}, and relax the volume and ionic positions within spin-polarized DFT. 
The   lattice constants optimized in this way for DMAMnF are found to be $a = 14.632$\,\AA, $b = 8.490$\,\AA, and $c = 9.057$\,\AA, with angles $\beta=59.3\degree$, and $\alpha=\gamma=90\degree$. This structure is the high-spin ambient-pressure room-temperature crystal structure. 

\begin{figure}
    \centering
    \includegraphics[width=\columnwidth]{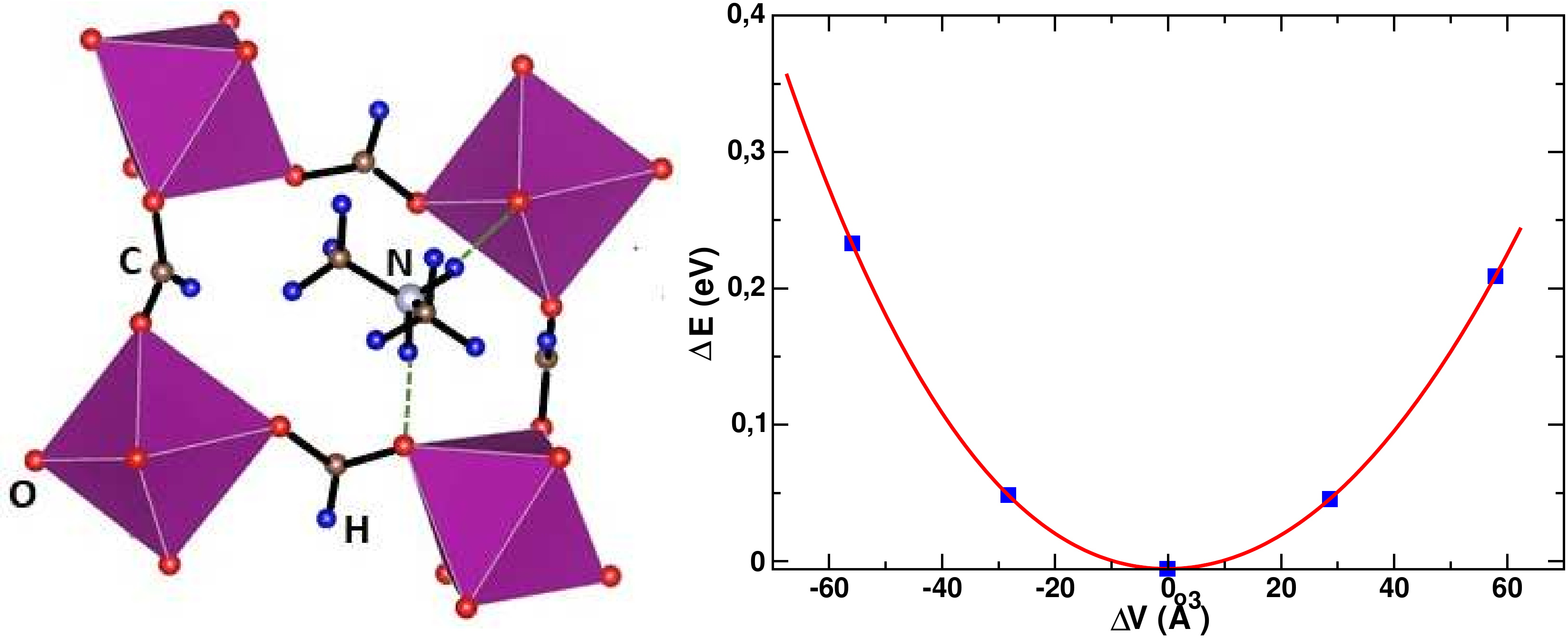}
    \caption{Structural details of DMAMnF. The left panel shows an image of the structure. The violet octahedra are the MnO$_6$ octahedra, with Mn being the B cation, connected by formate bridges forming the X anions, and the DMA cation sits in the cavity forming the A cation in the hybrid perovskite. The possible H bonds are marked by green dashed lines. The right panel shows a Birch-Murnaghan fit to obtain the bulk modulus of the material.}
    \label{struct}
\end{figure}
As shown in Figure \ref{struct}, in this compound each MnO$_6$ octahedron is connected to neighboring MnO$_6$ octahedra via HCOO$^-$ ligand bridges. This forms a three-dimensional ReO$_3$-type network with dimethylammonium cations occupying the centers of the ReO$_3$-type cavities. In DMAMnF, two bridging N-H···O hydrogen bonds from each DMA cation are formed. The strength of this hydrogen bonding is not very high due to the less polar nature of N-H bond as compared to O-H bonds. In Figure \ref{struct} we also show in the right panel the change in energy as function of volume change, which is fitted to the third-order Birch-Murnaghan equation. As discussed in the previous section, the bulk modulus is estimated from this fit to be 21.85\,GPa. This property can easily be measured experimentally. We also show the calculation and fit of the elastic properties, however, we discuss this in detail in a later section.

\subsection{Temperature and Pressure driven Spin Transition}

In this section we discuss in detail the temperature and pressure driven spin transitions in DMAMnF. 
The basic mechanism at work is simply that the HS state is realized for larger volumes and bond lengths, with a transition to a LS state for smaller volumes and bonds lengths. We found the critical bond length to be 1.95\,\AA. This spin transition as function of bond-length is also found in other materials~\cite{Jeschke_2007}. 
The transition from large to small volumes can now be either obtained by applying pressure, or by lowering the temperature. What we show in this section is that both types of transitions fall in parameter ranges that are easily accessible in experiments. 

\begin{figure}
    \centering
    \includegraphics[width=\columnwidth]{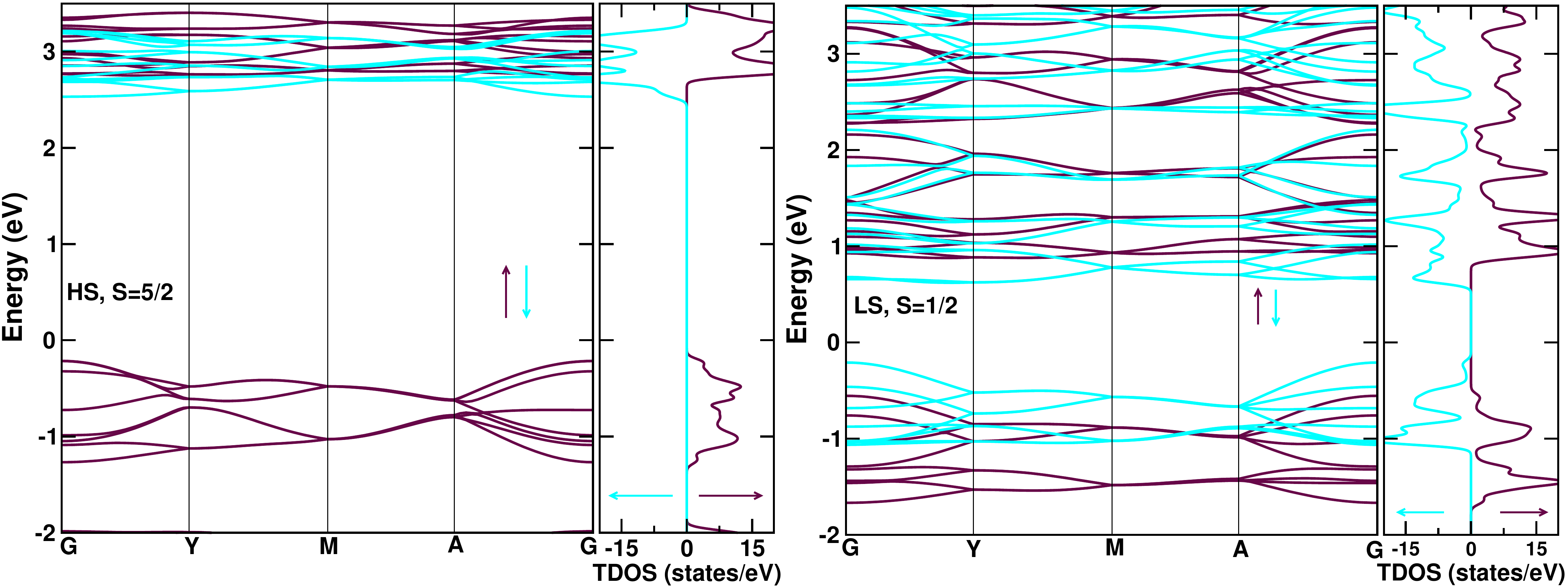}
    \caption{Figure showing the electronic structure of DMAMnF in HS and LS states. The left panel shows the spin-resolved band structure and DOS for the HS state, whereas the right panel shows the spin-resolved band structure and DOS for the LS state.}
    \label{electronic}
\end{figure}

We start with the pressure-driven transition. To get a starting point we carry out spin-polarised electronic structure calculation for the structure at ambient pressure as determined in the previous section.
The octahedral coordination of oxygen atoms around Mn splits the Mn $d$ states into states of $e_g$ and $t_{2g}$ symmetries. The Mn $d$ states are found to be completely occupied in the up-spin channel, and completely empty in the down-spin channel, giving rise to a perfect half-filling condition. An insulating ground state is obtained at ambient condition, with a large band gap ($\sim$ 2.5 eV). This suggests at ambient condition that the spin-state of Mn in DMAMnF is HS. The spin-resolved band structure and DOS are shown in the left panel of Figure \ref{electronic}. The calculated total magnetic moment (M) turned out to be 5\,$\mu_B$ per Mn atom, in agreement with the stabilization of the HS $(S = 5/2)$ state of Mn. 

\begin{figure}
    \centering
    \includegraphics[width=\columnwidth]{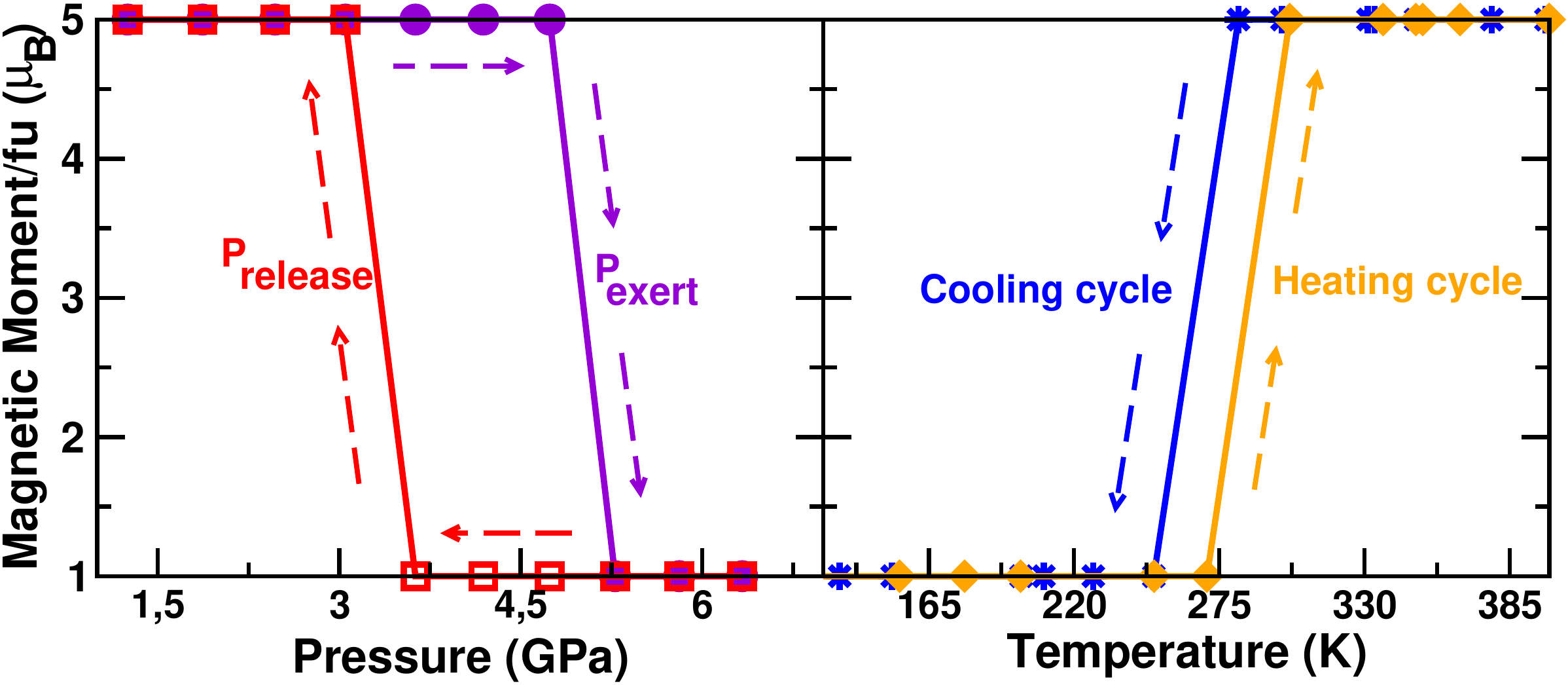}
    \caption{Pressure and temperature driven spin transitions accompanied by hystereses. The left panel shows the pressure-driven transition, whereas the right panel shows the temperature-driven transition obtained by a  combination of MD simulations.}
    \label{transition}
\end{figure}

Now we apply hydrostatic pressure to this ambient pressure HS structure. In order to find out the critical pressure where a spin-state transition happens, we increased the pressure in steps of 0.6-0.7\,GPa, starting from 0\,GPa.
As shown in the plot of the magnetic moment as function of pressure in Figure \ref{transition}, we find a spin-state transition from HS with total magnetic moment of 5\,$\mu_B$/Mn to LS with a total moment of 1\,$\mu_B$/Mn. The transition happens at a pressure $P_{c\uparrow}$ of 5.3\,GPa.

In the high-pressure LS state, we see a much smaller band gap of only $\sim 0.75$\,eV. The spin-resolved band structure for this state along with the DOS are shown in top right panel of figure \ref{electronic}. The spin-state transition is, thus, accompanied by a significant change in the overall band gap, which should be manifested in a corresponding change in optical response. We shall discuss this later in terms of optical absorption spectra and piezochromism.

We then released the pressure again, starting from the highest applied pressure. Of course, we find a transition from the LS state with a magnetic moment of 1\,$\mu_B$ back to the HS state with 5\,$\mu_B$ that is realized at ambient pressure. But interestingly, the pressure, where this transition happens, is $P_{c\downarrow}\approx 3.6$\,GPa, which is significantly lower than $P_{c\uparrow}$.
Thus we have a significant hysteresis effect in $M-P$ data, with a width of this hysteresis of 1.7\,GPa. We note that both $P_{c\uparrow}$ and $P_{c\downarrow}$ are of moderate values, which can be easily generated in a laboratory setup, and does not compromise the structural stability of the material. 

Next we investigate the temperature-driven spin transition. We turn to ab-initio molecular dynamics simulations at this step, to simulate the effect of temperature. Since we know that at low temperatures and at some applied pressure the system has smaller bond length and volume, we take as starting point a LS structure just above $P_{c\uparrow}$, and start with heating up the system as detailed in the Methods section. 

As shown in Figure \ref{transition}, we find a transition from a LS state with moment of 1$\mu_B$/Mn to a HS state of 5$\mu_B$/Mn during the heating cycle at a T$_c$ $\sim$ 300\,K. This brings the spin transition effectively close to room temperature, which has been one of most desirable features in spin-crossover materials. Most spin-crossover polymers until now have shown transitions at much lower temperatures. A room-temperature spin transition is a novelty which is yet to be seen in spin-crossover materials in general, and hybrid perovskites in particular. 

After increasing the temperature to 450\,K, we start decreasing it again, to explore the possibility of a hysteresis as function of temperature. We find that a transition from a HS state of 5$\mu_B$/Mn to 1$\mu_B$/Mn occurs at a different T$_c$ $\sim$ 250K. Thus, this shows a very clear hysteresis width of 50\,K, centred around 275\,K, which is a rather large hysteresis width at a temperature very close to room temperature. \HB{It may be noted that this AIMD calculations based on DFT are eventually mean field calculations which may involve some overstimation of critical temperatures.}

\MA{A few words are still in order concerning the order-disorder transition in these compounds. 
Experimentally it is known that DMAMnF shows 
a order-disorder phase transition at a critical temperature  $T_C=190$\,K . However, it is not possible in DFT studies to spontaneously change crystal structures, and go from an ordered to a disordered state by simple AIMD calculations. However, it has to be stressed that the disorder is observed in the A-site cation, and not in the transition metal octahedra. Since our main finding is that the spin state is primarily determined by the Mn-O bondlengths in those MnO$_6$ octahedra, we can savely assume that this effect is the same in both the ordered and disordered phases. As a result, the order-disorder critical temperature of $T_C=190$\,K is  unrelated to the calculated critical temperature is for a spin transition. There is absolutely no rationale that the two temperatures have to coincide. Both, HS and LS states can exist on both ordered and disordered structures.
Moreover a HS state can exist in a disordered state and a LS state can exist in an ordered state.}



Both the critical temperature and pressure \MA{for the spin transition, along with the width of hysteresis, may be tuned 
by changing the A-site amine cation, which contributes significantly to the hydrogen bonding and, hence, to the elastic properties of the system. Another possibility to reach pressures that are relevant for real device applications would be to replace partly the needed external pressure by internal chemical pressure, by going to larger A-site cations.} We shall discuss this in the subsequent section where we try to understand the microscopic origin of the cooperativity and associated hysteresis in hybrid perovskites. 

To sum up this section, we have demonstrated that spin transitions associated with large hysteresis widths is possible in a experimentally available hybrid perovskite, at reasonable laboratory conditions.

\subsection{Spin dependent Piezochromism}
In this section we discuss the possibility of 
piezochromism in this class of hybrid perovskites, a property which may be extremely useful in the design of display devices, photovoltaics or optoelectronic devices. Although certain Pb-halide based materials \cite{Umeyama2016} as well as some lead-free halides \cite{Jaffe2015, arnab-piezo} have been shown to demonstrate piezochromism, to the best of our knowledge there has been no suggestion of such behaviour in this class of formate hybrid pervoskites.
\begin{figure}
    \centering
    \includegraphics[width=\columnwidth]{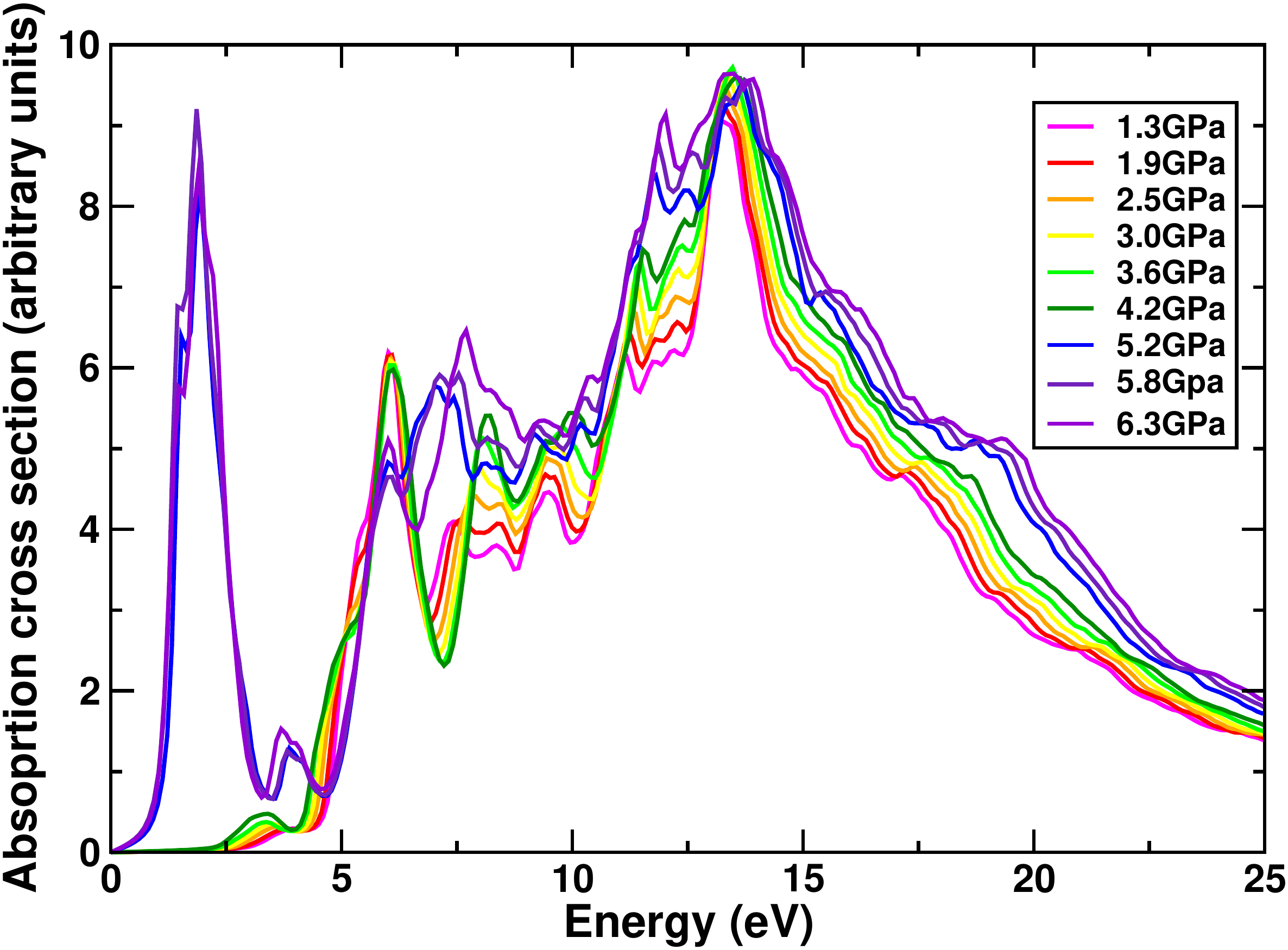}
    \caption{Optical absorption spectra for DMAMnF across the pressure-driven spin transition. The disappearance of the prominent peak around 2\,eV absorption energy in the HS state is clearly visible.  
    }
    \label{optical}
\end{figure}
 
 The pressure dependence of the absorption cross-section of the chalcogenide perovskites is depicted in Figure \ref{optical} at selected pressures from 0 to 6\,GPa. It is clearly visible that the cross sections are very different, in particular below roughly 5\,eV absorption energy.
 For the HS state at low pressures, we see a gap of roughly 2.5\,eV in the spectra, and a first strong maximum only above 5\,eV. For the LS state at high pressures, however, we see a strong peak at much lower energies of about 2\,eV
 The differences in the absorption spectra, in particular the different gaps, match well with the band gaps reported in figure~\ref{electronic}. In both  the HS and LS states, several peaks in absorption cross sections are observed. These additional absorption peaks indicate the multivalley characteristic of band structures, which primarily originate due to the allowed optical transition between primary and secondary valence band maxima and conduction band minima of the studied systems. On application of pressure on the HS state structure, a small red shift is seen, which may be attributed to the decrements of the bandgap in the optical absorption spectra. Eventually at a certain pressure we see a large change in the optical spectra which correspond to the transition to a LS state. 
 
 This big change in the absorption cross section also indicates a significant change in the color of the material, although it is difficult to predict the precise colour without doing excited state calculations. A simple hand-waving argument, looking at the energies where the absorption spectra become finite, could indicate a change in color from yellow to black. Similar color changes with applied pressure from red to black in case of Pb halide based halide perovskite~\cite{Umeyama2016}, and yellow to black in case of Cu halide based hybrid perovskite \cite{Jaffe2015} have been noted in literature. However, the transitions on those materials have not been associated with any spin transition.
 
Although the change in colour in different spin states is not unheard of in other spin-crossover polymers, this has, however, not been reported before in hybrid perovskites. In addition, the external stimuli pressure or temperature, which are necessary to induce this transition, are rather moderate. This should make this effect of piezochromism accessible in experiments.




 
\subsection{Microscopic understanding}
\begin{figure}
    \centering
    \includegraphics[width=0.75\columnwidth]{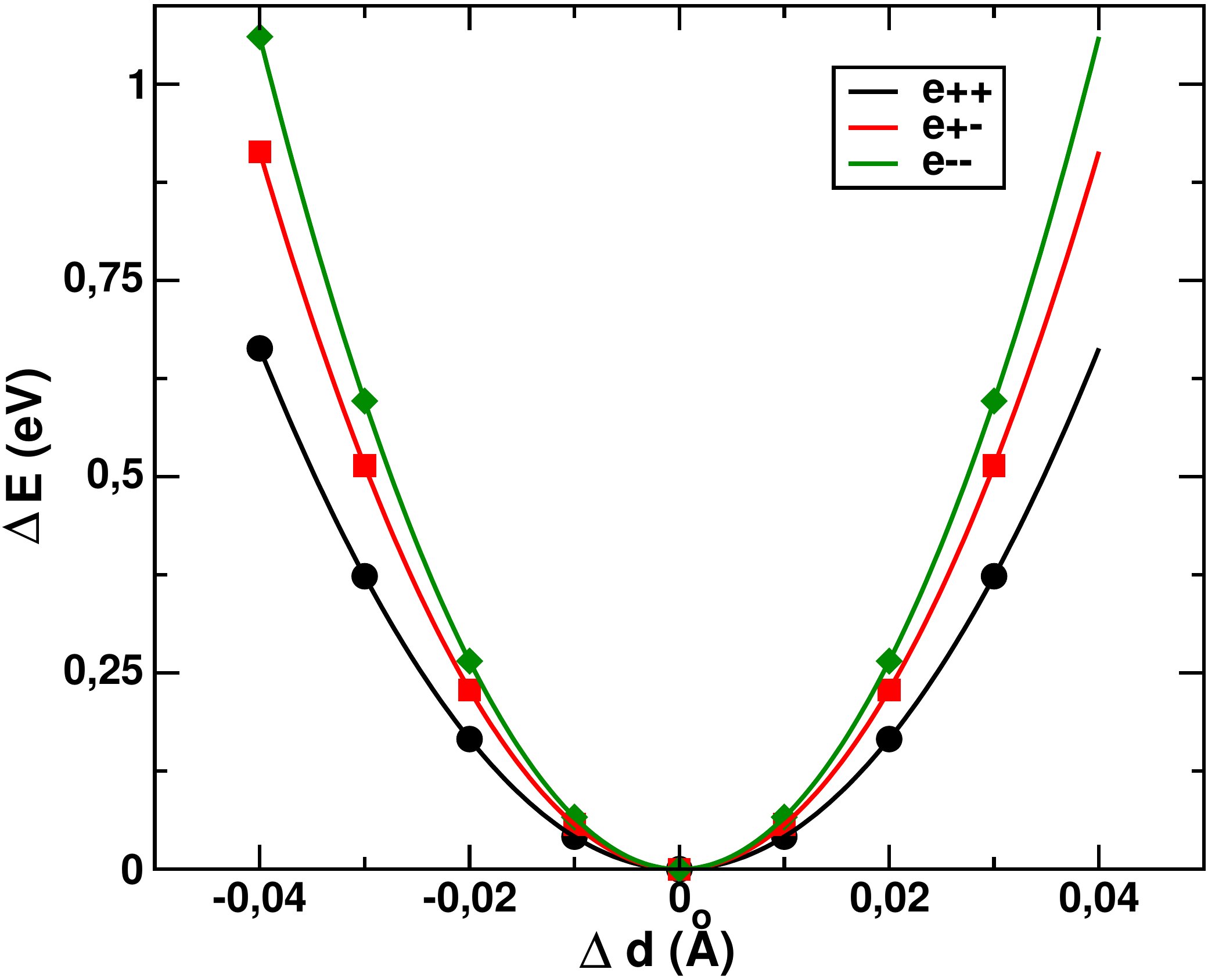}
    \caption{The figure shows the fit of the plots of Change in energy to change in bond-length to calculate the elastic constants.}
    \label{elastic}
\end{figure}
  
To explain the mechanism by which the cooperativity, manifested in the form of hysteresis in the $M$-$P$, and $M$-$T$ plots, emerges in this material we computed the elastic and magnetic exchange interactions. The most prevalent idea in this context attributes the microscopic origin of cooperativity to the elastic interaction between local distortions at the SCO centers \cite{Boukheddaden}. However, in a very recent work, the importance of magnetic superexchanges in driving cooperativity was also highlighted \cite{Banerjee2014}. Depending on the sign of the spin-dependent elastic interaction, which is dictated by the nature of the spin-lattice coupling in the material, the magnetic interaction was found to influence the hysteresis in a quantitative or qualitative way. The interplay between elastic and magnetic interactions in building up cooperativity crucially relies on the spin-dependent rigidity of the lattice. 
The elastic interaction between two neighboring Mn ions can depend on their spin state. These interactions are labeled $e_{++}$ ($e_{--}$) for both ions in HS (LS) state, and $e_{+-}$ for one site in HS and another in LS state. 

%
 The size change of the octahedral spin transition unit upon change of the spin-state makes $e_{++} < e_{--}$ \cite{Boukheddaden, Banerjee2014}. It is, thus, the value of $e_{+-}$ that decides the nature ( and sign) of the effective elastic interaction \cite{Boukheddaden, Banerjee2014}. This effective elastic interaction turns out to be of ferroelastic nature for $e_{+-} > \sqrt{e_{++} \times e_{--}}$, and of antiferroelastic nature for $e_{+-} < \sqrt{e_{++} \times e_{--}}$. It was demonstrated \cite{Banerjee2014} that for ferroelastic interactions, the magnetic interaction becomes operative only in a quantitative manner, in terms of enhancing the hysteresis width, while for antiferro elastic interactions, the magnetic exchange is the sole driving force in setting up the hysteresis.

To gain an understanding of the interplay of elastic and magnetic interactions in this case, we first calculated the spin-state-dependent elastic interactions. 
From our pressure studies we know that the LS state is obtained at higher pressures with an average Mn-O bond length of 1.9\,\AA{} or less, while the HS state is obtained at ambient pressures with an average Mn-O bond length of 2.2\,\AA.
Keeping this in mind, we constructed crystal structures setting the average Mn-O bond length at 1.9\,\AA{} for the LS structure and 2.2\,\AA{} for the HS structure. 
To simulate the HS-LS situation, a structure with alternating arrangements of Mn-O$_6$ octahedra having average Mn-O bond lengths of 1.9 and 2.2\,\AA{} was constructed. Considering the three model structures with HS-HS, LS-LS, and HS-LS arrangements of neighboring Mn-O$_6$ octahedra, the Mn-O bond lengths were varied by small amounts ($\sim 0.01–0.04$\,\AA), which is within the harmonic oscillation limit around the equilibrium bond lengths. The obtained energy versus bond-length variation for the three cases is shown in Figure \ref{elastic}. A parabolic fit of the data points provides the estimates of the spin-dependent elastic interactions. 

DMAMnF is found to be ferroelastic, with $e_{+-}>\sqrt{e_{++} \times e_{--}}$.
The values of elastic constants from the fitted curves are
$e_{++}=414.6$\,eV/\AA$^2$, 
$e_{+-}=571.2$\,eV/\AA$^2$, and
$e_{--}=662.7$\,eV/\AA$^2$.
These values may be obtained experimentally from calorimetric measurements. The strong ferroelasticity in the material may be attributed to the hydrogen bonds between N-H in the DMA cation and oxygen atoms in the MnO$_6$ octahedra. 

We next calculate the magnetic super-exchange coupling between neighboring Mn(II) centers in both HS and LS state. To estimate their values we calculated the total energies of ferromagnetic and antiferromagnetic Mn$^{2+}$ spin configurations and mapped them to  a model spin Hamiltonian, $H=\sum_{\langle ij\rangle}J_{ij}S_iS_j$, where $J_{ij}$ is the magnetic exchange between nearest neighbor Mn$^{2+}$ spins along a bond $\langle ij\rangle$. Due to the crystal structure, we consider two different exchange paths with constants $J_1$ in plane and $J_2$ out of plane. 
The difference of the ferromagnetic and antiferromagnetic energies provides an estimate of $J_{ij}$. Doing so, we obtain antiferromagnetic exchange in all cases, with coupling constants $J_1=7.43$\,K and $J_2=6.81$\,K for the HS state, as well as $J_1=4.63$\,K and $J_2=6.09$\,K for the LS state.


Following the previous literature \cite{Boukheddaden, Banerjee2014} we also conclude that the primarily responsible factor in driving cooperativity in these Mn-formate frameworks is the spin-dependent lattice effect, which is only enhanced by the magnetic super-exchange. A possible method of tuning the cooperativity, i.e., the width of hysteresis in this material, may be by changing the A-site cation, thereby changing the elastic constants of the material as has been demonstrated in a previous work \cite{Banerjee2016}.

\section{Conclusion}
In this article we have predicted the emergence of two new functionalities in a formate hybrid perovskite material of recent interest. We have shown that both pressure and temperature driven spin transitions, accompanied by hysteresis, are possible in DMaMnF, a Mn-based formate hybrid perovskite. 
Our study shows that the critical pressure is easily accessible in laboratory conditions, with a wide usable hysteresis width. 
Our three-step molecular dynamics calculations demonstrate that a room temperature spin transition along with a large width hysteresis is also possible in these materials, thus \MA{could make} it very convenient for usage in memory devices. We also indicate the possibility of piezochromism in these materials, which is particularly important for optoelectronic devices, and in photovoltaics, and has numerous possible applications. Summing up, our study adds another new dimension to the already wide field of possible applications of hybrid perovskites. 

\begin{acknowledgments}
The authors acknowledge funding support by the Austrian Science Fund (FWF): Y746. The calculations were performed on the Vienna Scientific Cluster and the dcluster of TU Graz. HB gratefully acknowledges useful discussions with Dr. Sudip Chakraborty.
\end{acknowledgments}



\bibliography{main} 

\end{document}